%% file: main.tex
\documentclass[journal]{vgtc}                

\ifpdf
  \pdfoutput=1\relax                   
  \usepackage{graphicx}                
  \DeclareGraphicsExtensions{.pdf,.png,.jpg,.jpeg} 
\else
  \usepackage{graphicx}                
  \DeclareGraphicsExtensions{.eps}     
\fi%
\usepackage{float}
\graphicspath{{figures/}{images/}{./}} 
\usepackage{microtype}                 
\PassOptionsToPackage{warn}{textcomp}  
\usepackage{textcomp}                  
\usepackage{mathptmx}                  
\usepackage{times}                     
\usepackage{cite}                      
\usepackage{tabu}                      
\usepackage{booktabs}                  
\usepackage{amsmath,amsfonts}
\usepackage{xcolor}
\DeclareMathOperator*{\argmax}{argmax}
\usepackage{algorithm}  
\usepackage{algpseudocode}
\usepackage{setspace}

\onlineid{1016}

\vgtccategory{Research}
\vgtcpapertype{algorithm/technique}

\title{Task-Oriented Optimal Sequencing of Visualization Charts}

\author{Danqing Shi, Yang Shi, Xinyue Xu, Nan Chen, Siwei Fu, Hongjin Wu, Nan Cao}
\authorfooter{
\item
 Danqing Shi, Yang Shi, Xinyue Xu, Nan Chen, Siwei Fu, Hongjin Wu, and Nan Cao are with Tongji University. E-mail: nan.cao@tongji.edu.cn.
}

\abstract{
A chart sequence is used to describe a series of visualization charts generated in the exploratory analysis by data analysts.
It provides information details in each chart as well as a logical relationship among charts.
While existing research targets on generating chart sequences that match human's perceptions, little attention has been paid to formulate task-oriented connections between charts in a chart design space. We present a novel chart sequencing method based on reinforcement learning to capture the connections between charts in the context of three major analysis tasks, including correlation analysis, anomaly detection, and cluster analysis. The proposed method formulates a chart sequencing procedure as an optimization problem, which seeks an optimal policy to sequencing charts for the specific analysis task. In our method, a novel reward function is introduced, which takes both the analysis task and the factor of human cognition into consideration. We conducted one case study and two user studies to evaluate the effectiveness of our method under the application scenarios of visualization demonstration, sequencing charts for reasoning analysis results, and making a chart design choice. The study results showed the power of our method. 

} 


\CCScatlist{ 
 \CCScat{K.6.1}{Management of Computing and Information Systems}%
{Project and People Management}{Life Cycle};
 \CCScat{K.7.m}{The Computing Profession}{Miscellaneous}{Ethics}
}

\teaser{
 \centering
 \includegraphics[width=\linewidth]{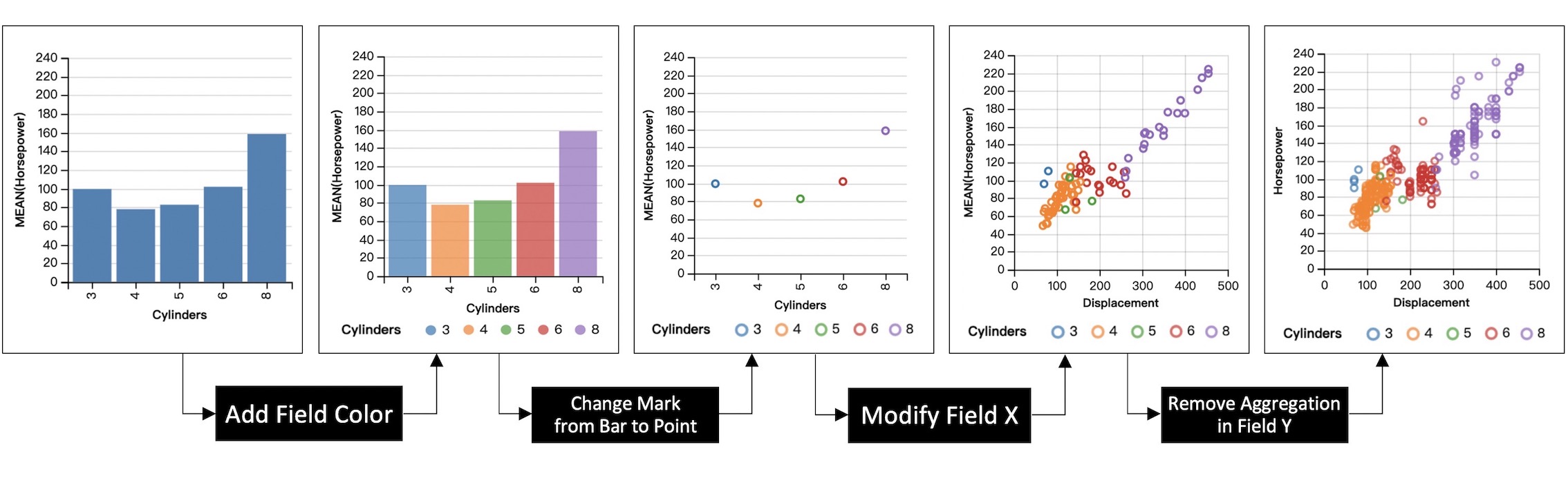}
 \caption{A case study based on the optimal task-oriented charts sequencing technique. This figure illustrates the results of an expert user exploring the Cars dataset starting from a bar chart (the first chart on the left) to find out the clusters of cars having the same number of cylinders. During the analysis, our technique recommends proper actions to guide the exploration process to approach the analysis goal. After a series of operations, the user finally found that cars are clustered in a subspace as shown in the last view, which indicates that fewer cylinders corresponds to smaller displacement and the lower horsepower.}
	\label{fig:teaser}
}

\renewcommand{\manuscriptnotetxt}{}

\vgtcinsertpkg

\begin{document}

\maketitle

\input{01-intro}

\input{02-relatedwork}
\input{03-background}
\input{04-reward}
\input{05-evaluation}
\input{06-discussion}
\input{07-conclusion}

\acknowledgments{
We would like to thank all visualization experts and users for participating our studies and interviews. This research was supported in part by NSFC Grants 61602306, Fundamental Research Funds for the Central Universities, and the National Grants for the Thousand Young Talents in China.
}

\bibliographystyle{abbrv-doi}

\bibliography{template}
\end{document}

%% file: 01-intro.tex
\section{Introduction} 

A chart sequence is used to describe a series of visualization charts generated in the exploratory analysis by data analysts. 
It serves as a form of graphical history which help analysts review their prior findings\cite{Heer2008GraphicalHF} or a tour guide which helps explore complex datasets\cite{TheGrandTour}. 
Chart sequences not only provide information details in each chart, but more importantly, they assist interpreting the exploratory process, helping analysts understand the data, and making the following decisions~\cite{Kim2017GraphScapeAM}. 
For example, in an anomaly detection task, a user not only wants to investigate the anomaly but also analyze how the anomaly is emerged from the data. An effective chart sequence can help illustrate the entire analysis process and interpret how an anomaly is detected. 

Due to the importance, recently, research attentions have been put on developing techniques to threading visualization charts into meaningful sequences. Kim et~al.~\cite{Kim2017GraphScapeAM} introduced GraphScape, the state-of-the-art technique, that uses a directed graph to model the entire chart design space with the nodes indicating design states (i.e., a chart with proper parameters and data mapping) and the links indicating various design actions such as changing the chart type or applying a new data mapping. A user study was conducted to estimate each of the design actions with the goal of weighting links (i.e., actions) in the graph based on their capability of preserving a user's perception. As a result, a context-preserving chart sequence corresponds to a path with the maximum weight in the graph. 
Although GraphScape can suggest efficient sequence in general case, it still has limitations in two scenarios: (1) Multiple chart sequences can interpret the transition from the start chart to target chart with the same perception cost. Thus analysts still have to take effort to pick sequence;
(2) Choosing different chart sequences may affect the subsequent decision based on the specific analysis task. Therefore, only considering user perception is not enough.

To address the above issues, in this paper, we introduce a novel chart sequencing method based on the chart design space and the graph model introduced in GraphScape. 
Our method can recommend a sequence of visualization charts to help a user to travel from the current visualization to the desired visualization based on the given task. 
Our technique estimates the weight of the design actions through a reinforcement learning based approach by considering three common tasks in visual analysis: (1) correlation analysis; (2) anomaly detection, and (3) cluster analysis. In our methods, we make an analogy between a Markov decision process (MDP) and a chart sequencing procedure where the identification of chart sequences can be framed as finding an optimal decision policy through the MDP. In particular, given a chart design space represented by a directed graph and modeled by MDP, we aim to find an optimal path connecting a set of charts that best matches the analytical process of a specific task. We introduce an inverse reinforcement learning technique to learn the reward score of each action via a small number of analysis demonstrations performed by experts. As a result, we obtain a reward function that incorporates both analysis tasks and human cognition factors. Finally we apply a value-iteration based reinforcement learning algorithm based on the rewards to find the optimal policy to achieve different tasks. Our approach supports sequencing visualization charts for reasoning an analysis result and offering a chart design choice for decision making. We evaluate the proposed technique through one case study and two user studies to estimate its capability of supporting  visualization recommendation, reasoning analysis results, and making chart design choice.  Generally, the paper has the following contributions: 

\begin{itemize}
    \item 
    We model a chart design space as a Markov decision process, and 
    propose an approach based on reinforcement learning that seeks an optimal policy to sequencing charts in the design space to achieve a specific analysis task. 

    \item We propose an inverse reinforcement learning method to learn a reward function that takes into account both the analysis tasks and human cognition. 

    \item We conduct a case study and two controlled user studies to evaluate the effectiveness of our approach in the application of visualization recommendation, sequencing charts for reasoning an analysis result, and making a chart design choice. 
\end{itemize}

The rest of this paper is structured as follows. We describe related research work in Section 2. In Section 3, We present the reinforcement learning based chart sequencing approach in detail. Section 4 introduces one case study and two user studies to evaluate our technique, from which we discuss the limitations and implications in Section 5. Finally, we summarize our work and future research directions in Section 6.

%% file: 02-relatedwork.tex
\section{Related Work}
In this section, we review related techniques in four categories: (1) visualization sequence, (2) visualization state space, (3) visualization recommendation systems, and (4) task-based evaluation.

\subsection{Visualization Sequence}

Users usually explore visualizations in an interactive way for visual analysis\cite{Heer:2012}.
The whole exploratory analysis process can be presented as a visualization sequence.
Graphic history interface recording a sequence of visualizations helps analysts to review, retrieve, and revisit their prior findings\cite{Heer2008GraphicalHF}.
Insight provenance can be derived from historical records of user exploration and analysis process.
Gotz and Zhou characterize users’ visual analytic activity at multiple levels of granularity, i.e., Task, Sub-Task, Actions, and Events\cite{Gotz2008CharacterizingUV}. 
They found that Actions can be used to represent activity both generally and semantically. 
Later, they investigated patterns from recorded sequences and observed the four most widespread behavior across different users and tasks, such as Scan, Flip, Swap, and Drill-Down\cite{Gotz2009BehaviordrivenVR}. 
Their research can be used to recommend visualizations according to user interaction.
Similarly, Bavoil et~al. proposed VisTrails that provides an infrastructure to streamline the creation and execution of visualization exploration pipelines\cite{Bavoil2005VisTrailsEI}. 
They further track the workflows to leverage provenance information to automate the construction of new visualizations\cite{Callahan2006ManagingTE, Scheidegger2007QueryingAC}.

Visualization sequences are also commonly seen in narrative visualization.
According to Segel and Heer\cite{Segel2010NarrativeVT}, narrative flow often seeks to produce balances between author-driven and reader-driven experiences.
Some narrative flow is crafted according to the intention of authors, which focuses on the logic of telling a goal-based story.
Some authors may employ animated transition to make the presentation more attractive\cite{Heer2007AnimatedTI}.
Authors often organize visualizations in a linear sequence structure when they provide a presentation or write a report\cite{Segel2010NarrativeVT}. 
Hullman et~al.\cite{Hullman2013ADU} showed the evidence from cognitive psychology that the sequence structure of linear-style narrative visualization plays a vital role in effective storytelling. 
Later, they found that hierarchical sequence structure characterizes most preferred visualization sequences\cite{Hullman2017FindingAC}. 
Recently, Qu and Hullman contributed detailed characterization of authors’ rationales for tolerating inconsistencies under some conditions\cite{Qu2018KeepingMV}. 

Our approach draws inspiration from aforementioned research. 
We extend prior work by taking analysis tasks into consideration.
We model a chart sequencing procedure as a Markov decision process, and propose an inverse reinforcement learning method to learn a reward function that reflects both analytical process and human cognition.

\subsection{Visualization State Spaces}

Visualization state space can be fit in a unified graph-based model where individual visualizations are nodes in the graph; 
users can trace a path through the graph as they explore visualizations. 
Image Graph is the first graph-structure representation for visualization exploration\cite{Ma1999ImageGN}. 
Each visualization image and its parameters are modeled as a node while the parameters change between two nodes are links.
P-Set model\cite{JankunKelly2007AMA} extends Image Graph by introducing a framework to encapsulate, share, and analyze the process of visual exploration.
The model can be used to operate upon or analyze a wide domain of visualization in a rigorous manner.

Hullman et al. informed a graph-based approach that identifies possible sequences in a visualization set and a visualization-to-visualization transition cost model that approximates the cognitive cost\cite{Hullman2013ADU}.
Based on the concept, Kim et~al.\cite{Kim2017GraphScapeAM} proposed GraphScape, a directed graph model for reasoning about a visualization state space.
Nodes in the graph are Vega-Lite\cite{Satyanarayan2017VegaLiteAG} charts specifications, and edges are edit operations between two specifications. 
Though similar to our work, GraphScape has its limitation in two aspects.
First, it does not incorporate analysis tasks in modeling the visualization state space.
Thus, it is unable to answer questions like, how different sequences of visualizations affect the analytic process.
Second, GraphScape may identify and rank multiple paths according to human perception in the application of path elaboration. However, it unables to rank paths from the perspective of analysis tasks.
In this work, we address the above limitations by modeling the connections between charts in the context of three different analysis tasks, including cluster analysis, anomaly detection, and correlation analysis.

\subsection{Visualization Recommendation Systems}

Visualization recommendation systems often model a visualization state space and use an objective function to suggest subsequent charts according to the context, which is relevant to our work. 
There are two lines of research in this field, i.e., rule-based and learning-based. 
Rule-based approaches require experts to manually craft rules to guide visualization design.
APT is the first to automate the design of 2D graphical presentation for data using perceptual principles\cite{Mackinlay1986AutomatingTD}. 
ShowMe is an automation tool which is incorporated in Tableau to present data with support for dimension selection\cite{Mackinlay2007ShowMA}. 
Voyager extends prior work by automatic generation of a diverse set of visualizations\cite{Wongsuphasawat2016VoyagerEA}. 
Draco supports the design of visualizations by encapsulating design knowledge as constraints\cite{Moritz2018FormalizingVD}. 
Kim and Heer \cite{Kim2018AssessingEO} consider analysis tasks to recommend effective visual encoding for automated visualization design.
Although hand-crafted rules are necessary for the visualization research, our approach tries to learn inherent knowledge of decision making through the demonstration of the analytic process performed by visualization experts.

With rapid development of machine learning, learning-based approach is becoming increasingly popular in recent years. For example, 
VizML\cite{Hu2018VizMLAM} identifies five visualization design choices and trains a machine learning model to learn the design choices from a large corpus of datasets.
DeepEye\cite{Luo2018DeepEyeTA} trains a machine learning model with large datasets to find top-k visualizations for input data.
Data2Vis borrows a deep learning model that formulates visualization design as a sequence to sequence translation problem\cite{Dibia2018Data2VisAG}. 
The research mentioned above requires large-scale and high-quality training datasets, which are formulated as data-to-visualization or task-to-visualization pairs.
Recently, a new large-scale visualization dataset as a collection of (data, visualization, task) triplets \cite{Hu2018VizNet} is in construction.
In our inverse reinforcement learning method, we use the demonstrations of analytic activities by experienced visualization experts as our training data, which are easy to collect using the log files in visualization tools.

\begin{figure*}[!htb]
\setlength{\abovecaptionskip}{15pt}
\centering 
\includegraphics[width=2\columnwidth]{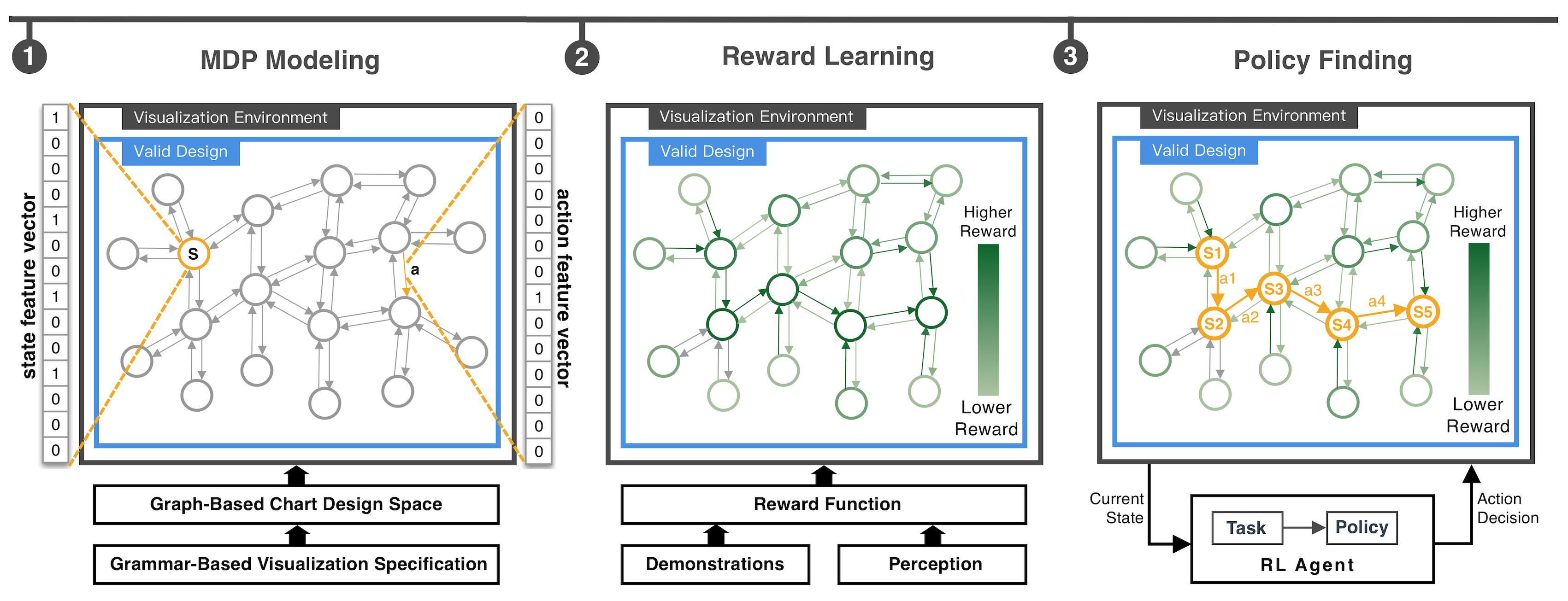}
\caption{The proposed technique for generating optimal task-oriented chart sequences, which consists of three major steps: (1)  design space modeling based on a Markov Decision Process (MDP); (2) reward learning based on the max entropy inverse reinforcement learning; (3) chart sequencing based on an optimal state policy in the design space defined by the MDP via reinforcement learning using the reward function learned in the last step.}
\label{fig:environment}
\end{figure*}

\subsection{Task-driven Evaluation}
Visualizations help viewers leverage visual system to achieve certain tasks\cite{SHGF16}.
A number of research have studied empirically the connections between visual encodings and different analysis tasks. For example,
Gleicher\cite{GCNF13} evaluated the human ability to compare average value in multiclass scatterplots.
Similarily, Albers\cite{ACG14} studied how visual encodings may support various aggregate comparison tasks.
Some research\cite{HYFC14,KH16} borrow Weber\rq s law to model the precision of estimation of correlation in different visualizations.
A  recent trend is the study of visual preference in achieving a broad range of analysis tasks. For example,
Saket et~al. \cite{Saket2018TaskBasedEO} conducted a crowdsourced experiment to evaluate the effectiveness of five types of visualization across the ten low-level tasks. 
Recently, Kim and Heer \cite{Kim2018AssessingEO} accessed the effectiveness of visual encodings based on analysis tasks being performed.
These approaches focus on single visualization or encodings of one visualization. 
On the contrary, our work evaluates the effectiveness of a sequence of visualizations in the context of analysis tasks. 

%% file: 03-background.tex
\section{Task-Oriented Sequencing of Charts}
In this section, we propose a reinforcement learning based technique to thread charts in a chart design space into perception-preserving sequences to approach a specific analysis goal, including correlation analysis, anomaly detection, and cluster analysis. 

\subsection{Overview}
\autoref{fig:environment} illustrates the overview of the proposed technique, which consists of three main steps: (1) We first model the chart design space as a Markov Decision Process (MDP) with reward functions left blank; Then, (2) we learn a reward function for each analysis task using the inverse reinforcement learning method through a small number of analysis demonstrations performed by expert users. Furthermore, we incorporate human cognition into the reward function to facilitate the understanding and reasoning of chart sequences with lower perception cost; Finally, (3) we use the reinforcement learning algorithm based on the reward functions to find an optimal policy in the MDP. The optimal policy can thread charts into meaningful and task-oriented sequences. In the rest of the section, we will introduce each of these key steps in detail.

\subsection{Modeling the Chart Design Space}
In this section, we first review the background of a Markov Decision Process (MDP), followed by a detailed description of using MDP to represent a chart design space.

\subsubsection{Background of MDP}

A Markov decision process (MDP) describes a sequential decision making procedure in a dynamic environment based on a state transition model, in which state transits to another state through various actions under a certain probability~\cite{sutton1998introduction}. Formally, an MDP is defined as a tuple  $<S, A, R, P, \gamma>$, where $S=\{s_1, s_2, \cdots , s_m\}$ is the state space; $A=\{a_1, a_2, \cdots , a_n\}$ is the action space; $R : S \times A \times S \rightarrow \mathbb{R}$ is a reward function that determines the benefits that one will get after transferring from one state to another by performing a certain action; $P:S\times A\times S\rightarrow [0,1]$ is the transition function that captures the probability of transit from one state to another by performing an action; $\gamma \in [0,1)$ is a discount factor. 

The objective of an MDP is to find an optimal state transition \textit{policy} $\pi^*$ that maximizes the expected accumulative rewards starting from any given state $s$ to approaching a specific goal (i.e., a desired state). The optimization process can be formally defined as:
\begin{equation}
\pi^* = \arg\max_{\pi}\mathbb{E}_{\pi}\left[\sum_{k=0}^{\infty} \gamma^{k} r_{(t+k)} | s_{t}=s\right]
  \label{eq:mdpgoal}
\end{equation}
where $\pi : S \times A \rightarrow [0, 1]$ is the policy function which indicates the probability of performing a specific action given a state; $\pi^*$ is the optimal policy and $\mathbb{E}_{\pi}$ is the expectation under policy $\pi$; $\gamma \in [0,1)$ is the discount factor which is used to penalize the future rewards; $t$ indicates the current time and $r_{(t+k)}$ indicates the reward at time $t+k$, which is given by
\begin{equation}
r_{(t+k)} = R\left(s_{(t+k)}, a_{(t+k+1)}, s_{(t+k+1)}\right)    
\end{equation}

\begin{figure}[!htb]
 \centering 
 \includegraphics[width=\columnwidth]{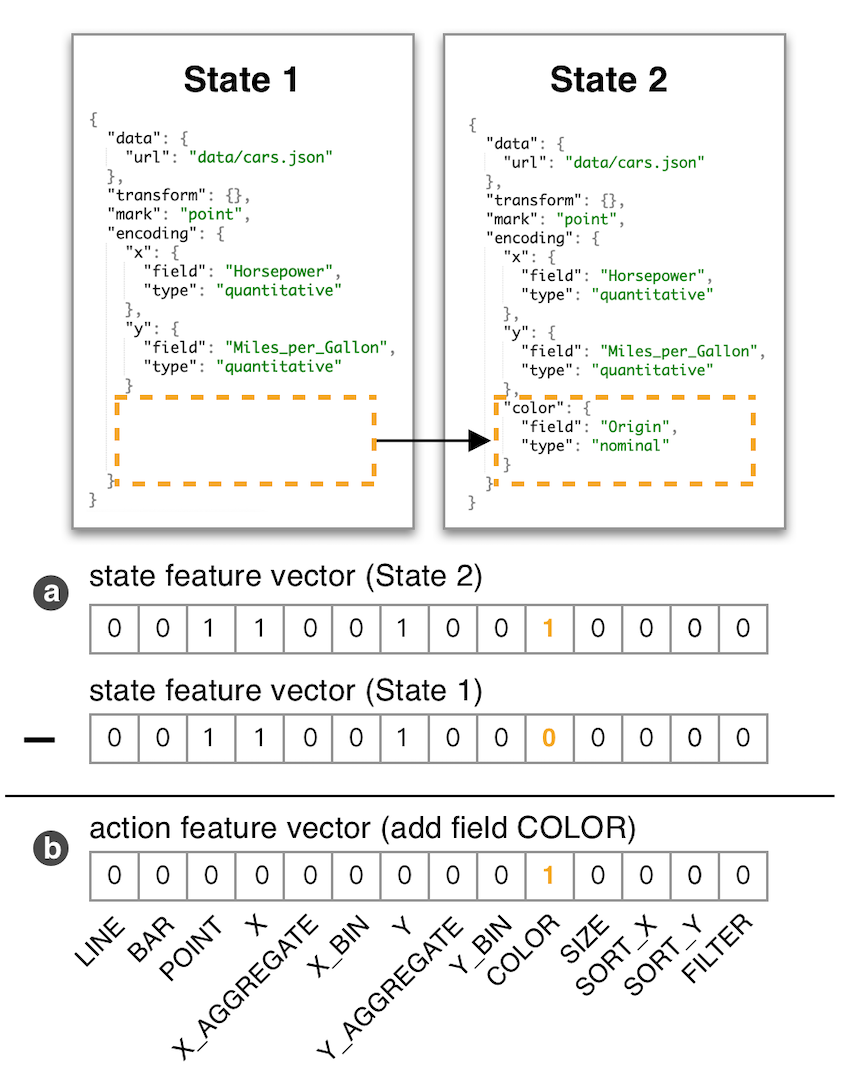}
 \caption{The subtraction of (a) two state feature vectors results in (b) an action feature vector.}
 \label{fig:states}
\end{figure}

\subsubsection{Design Space Modeling}
\label{subsec:modeling}

We adopt the graph model introduced in GraphScape~\cite{Kim2017GraphScapeAM} but extend it based on an MDP denoted as $\mathbf{M} = <S, A, R, P, \gamma>$ to capture the entire visualization chart design space. In particular, as introduced in GraphScape, a directed graph $\mathbf{G} = <V, E>$ is used to model the design space. Each graph node indicates a chart with properties (e.g., chart type) and proper data mappings and each edge indicates an editing behavior such as changing the chart type or encoding method. Even with a small number of charts and operations, the design space of chart sequence is already huge. In our implementation, we construct a design space with only three types of charts: bar chart, line chart, and scatter plot. We map this model to an MDP as follows:

\paragraph{\bf States ($S$)} We denote each node $n \in V$ in $G$ as a state $s \in S$ in $M$. To facilitate the calculation in an MDP, the states need to be further vectored.  To this end, we adopt the node representation in GraphScape, i.e., a grammar introduced in VegaLite~\cite{Satyanarayan2017VegaLiteAG}, which describes a chart by three descriptive components with 11 optional values (Fig.~\ref{fig:states}): (1) \textit{marks} with the possible options as ``bar", ``line", and ``point", (2) \textit{encoding} with possible options as ``x axis", ``y axis", ``color", and  ``size", and (3) \textit{transform} with possible options as ``aggregate", ``bin", ``sort", and ``filter". These options characterized a chart, which are used as the features to represent a state. Specifically, a 14-dimensional one-hot-vector $\mathbf{f}_{s}$ is introduced in our system to represent a state based on the combinations of these options as shown in Fig.~\ref{fig:states}. 

\paragraph{\bf Actions ($A$)} The action space $A$ is directly defined by $E$, i.e., the edge collections in $\mathbf{G}$. An action can also be represented by a feature vector which is derived from the state features. Intuitively, an edit operation is the reason causes the differences of two succeeding states in the design space, thus can be represented by the vector differences between two states (Fig.~\ref{fig:states}(b)): 
\begin{equation}
    \mathbf{f}_{a} = \mathbf{f}_{s_{t}} - \mathbf{f}_{s_{t - 1}}
\end{equation}

In addition, the action feature vectors of ``modify
    field x", ``modify field y", and ``modify field color" will be the same as ``add
    field x", ``add field y", and ``add field color". Because the actions change the same visual channel in each pair.

\paragraph{\bf Reward Function ($R$)} Directly build a reward function in our case is a challenging task. Although GraphScape successfully estimates a cost for each edit operation based on a user study, a reward function, $R:S \times A \times S \rightarrow \mathbb{R}$, takes the state space $S$ into consideration, thus resulting a much larger investigation space, which can hardly be manually estimated through a user study. For example, in our case, the design space $\mathbf{G}$, although only contains three types of charts, has 1152 nodes and 13056 edges, i.e. we need to estimate 13056 different transitions to build a reward function. In addition, when taking different analysis tasks into consideration, the problem becomes even harder. Furthermore, when considering different analysis tasks, we need to consider different analysis situations when sequencing the chart, thus making the problem becomes even harder. To address this issue, in this paper, we employed the technique of inverse reinforcement learning to build a task-oriented reward function through a series of analysis demonstrations made by a few expert users. The detailed techniques will be discussed in the next section.

\paragraph{\bf Transition Function ($P$)} In the design space $\mathbf{G}$, transition probability between two states is deterministic, i.e., either 1 (the two states are connected by an edge) or 0 (two states are disconnected). Therefore, the transition function can be defined as $P:S \times A \times S \rightarrow \{0,1\}$.

\paragraph{\bf Discount Factor ($\gamma$).} The discount factor determines the present value of future rewards. As $\gamma$ is close to 1, it will take future rewards into account very strongly. In our case of task-oriented chart sequencing, we set $\gamma = 0.99$.

Based on the above settings, a chart sequencing problem can be formulated as an optimal \textbf{policy} finding problem and solved based on a reinforcement learning algorithm.

%% file: 04-reward.tex
\subsection{Learning the Reward Function}
\label{sec:rl}
In this section, we introduce an inverse reinforcement learning (IRL) algorithm, the maximum entropy IRL~\cite{ziebart2008maximum}, that we employed to learning a reward function for both actions and states through a small set of training samples. We choose to use an IRL algorithm due to the lack of data and methods for directly estimating action rewards regarding to different states. Our training sample is a collection of visual analysis sequences generated by expert users during a data analysis process given three different tasks: (1) correlation analysis, (2) anomaly detection, and (3) cluster analysis.




\begin{figure*}[!ht]
\centering 
\includegraphics[width=0.8\linewidth]{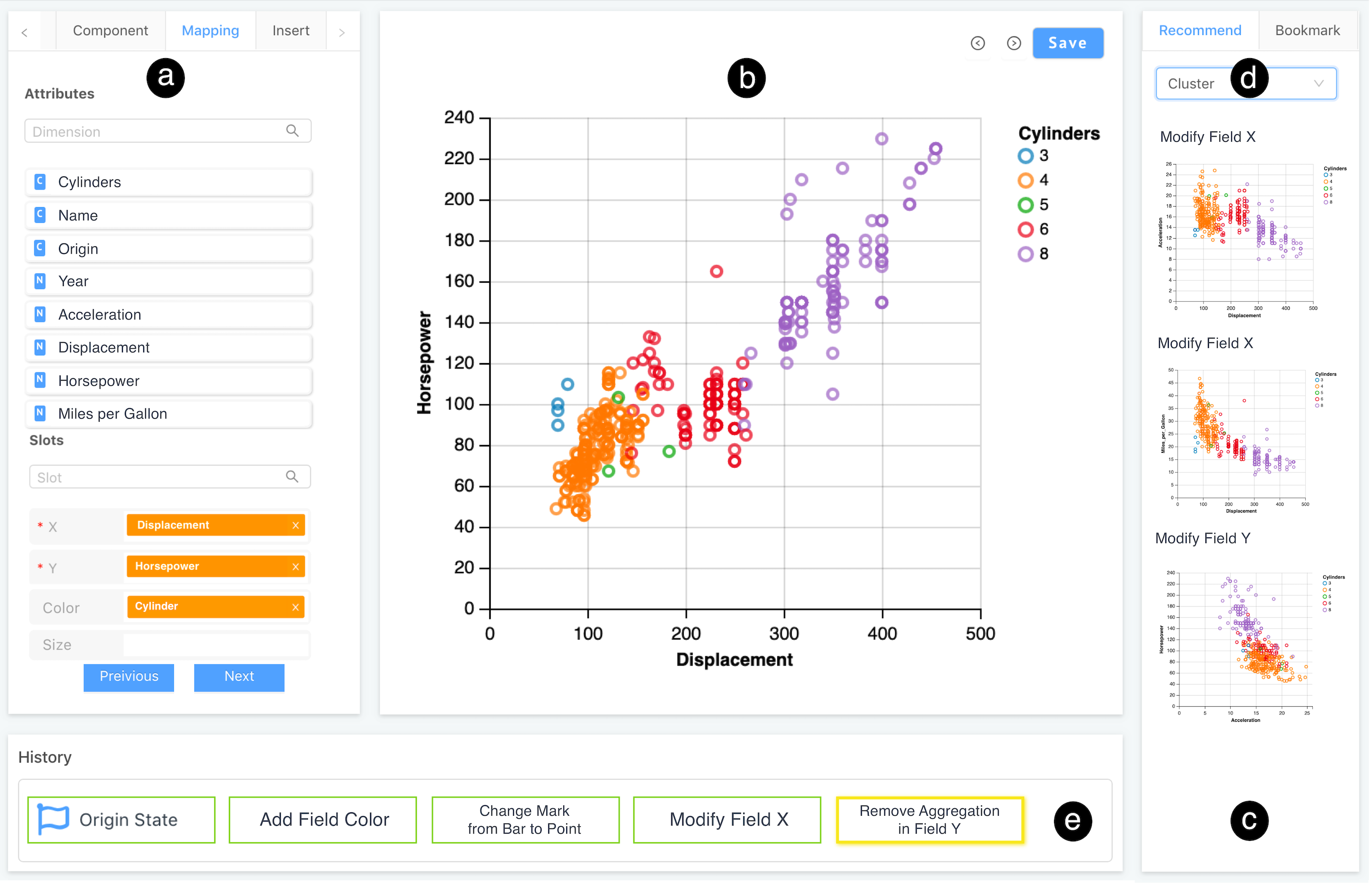}
\caption{A prototype system developed for evaluating the proposed chart sequencing technique, which consists of (a) a data mapping panel, (b) a chart editing panel, (c) a recommendation list, (d) an analysis task choicer, and (e) an action history list.}
\label{fig:systemui}
\end{figure*}

\subsubsection{Maximum Entropy IRL}
The maximum entropy IRL algorithm takes a small set of training sample, i.e., analysis sequences (which could be incomplete and biased), to learn a sequence distribution by following the principle of maximum entropy~\cite{maxent}, which helps to gain information from an incomplete and biased sample. The principle gives a reasonable strategy to obtain a distribution from small samples. In particular, given a small set of data samples, the data distribution can hardly be determined. Among various qualified distributions, the one with the maximum entropy is the most robust and reveals the most general situation regardless of the bias given by the small sample. 

With the above concept in mind, formally, the maximum entropy IRL is defined to maximize the entropy of the distribution of analysis sequences given a small training set based on the following objective: 
\begin{equation}
	\theta ^ { * } = \argmax_{\theta} \sum _  { i=1}^m  \log P ( \tilde{\zeta}_i | \mathbf{\theta})
  \label{eq:goal}
\end{equation}
where $\tilde{\zeta}_i$ is a training sample (i.e., an analysis sequence) generated by the experts; $\mathbf{\theta}$ is the parameter vector to be learned, which also directly defines the action and state rewards (introduced later); $P( \tilde{\zeta}_i | \mathbf{\theta} )$ shows the occurrence likelihood of a sequences in the training set under the parameters given by $\mathbf{\theta}$, which is defined as follows:
\begin{equation}
 P \left( \zeta_i | \theta \right) = \frac { 1 } { Z ( \theta ) } e ^ { \theta ^ { \top } \mathbf { f } _ { \zeta_i } }
 = \frac { 1 } { Z ( \theta ) } e ^ { \sum _{ s_ {j }, a_{j} \in \zeta_i } \theta ^ { \top } (\mathbf { f } _{ s_ { j } } + \mathbf { f } _{ a_ {j} }) }
 \label{eq:p}
\end{equation}
where $\mathbf{f}_{s_{i}}$ and $\mathbf{f}_{a_{i}}$ respectively indicate the feature vectors of a state $s_{i}$ and an action $a_{i}$, whose rewards are respectively defined by $\theta^{T}\mathbf{f}_{s_{i}}$ and $\theta^{T}\mathbf{f}_{a_{i}}$; $Z(\theta) =\sum_i e^{\theta^{T}\mathbf{f}_{\zeta_i}}$ normalizes the whole term into a probability. Following this definition, $\log P ( \tilde{\zeta}_i | \mathbf{\theta})$ indicates the entropy of the distribution captured by $P(\cdot)$. 

Please note that in Eq.~(\ref{eq:p}), we slightly changed the original definition of $\log P ( \tilde{\zeta}_i | \mathbf{\theta})$ by adding the term $\theta^{T}\mathbf{f}_{a_{i}}$ in the purpose of calculating an action reward based on the same algorithm framework.

\subsubsection{Reward Function}
Once trained, the above algorithm provides a reward function $R_T(\cdot)$ for each state and action in the chart design space $G$ modeled by an MDP $M$. Given the reward function, a chart sequence $\zeta_i$, starting from an arbitrary state $s_i$, can be generated to approach a given analysis task $T$. However, this sequence without taking a user's perception into consideration, thus may generate discontinuous sequences, thus greatly affects the readability of the chart sequencing results.

To address this problem, we borrow the perception-preserving costs, $c(a_i)$, of an edit operation $a_i$ (i.e., an action in our case) introduced in GraphScape as a part of the reward function. Formally, a reward function of a given task $T$ is defined as a linear combination of state reward, action reward, and $c(a_i)$ as follows:
\begin{equation}
	R_{T}(s_{(i-1)}, a_{i}, s_{i}) = (R_T(s_{(i-1)}) + R_T(a_{i})) + \lambda \cdot (-c(a_i))
\end{equation}
where the first two terms ensuring the given task $T$ can be achieved based on the reward and the last term penalizes the overall reward score based on the perception cost. $\lambda$ is a parameter that balances between these two parts, which is set to 0.3 in our implementation.

\subsubsection{Training the Model}

In our implementation, three reward functions were respectively trained for the aforementioned three analysis tasks, i.e., correlation analysis, anomaly detection, and cluster analysis. 

To collect training samples, we conducted a pilot study with expert users to collect their analysis sequences given a specific analysis task. Three expert users who were experienced with Tableau and visual analysis were recruited from an international business intelligent company. They were required to manually restoring an analysis sequence to approach a desired state (i.e., a chart with proper data mapping, attributes, and reveals a given data pattern) by fully exploring the entire chart design space defined by $\mathbf{G}$ and modeled by $\mathbf{M}$. 

Before the study, we prepared the experiment data and tasks by exploring the Tableau Public Gallery\cite{tableaupublic} and Plotly Community Feed\cite{plotlyfeed}. We filter the visualizations on these public platforms by the rules that the selected visualizations should have 100 views, one star, and one review at least. A set of 60 visualization views together with their data were collected from these sources with each of the views clearly shows an outlier, or a data correlation pattern, or data clusters, which correspond to the results of our focal analysis tasks. These views and data were latter used as study tasks, i.e., the ``desired state", in the study.

During the study, a user restored an analysis sequence starting from a random state in the design space and ending at one of the aforementioned views selected by us. We use random states as the initial state in each session to make sure the diversity of interactions. Each of the users was asked to took several minutes to understand the background of the data and then restore the analysis sequences for 20 selected visualization views by editing the initial chart in series based on the edit operations defined in $\mathbf{G}$. There were multiple ways to approach a desired view. Only the one that best preserved the users' cognition and with a shorter length (i.e., more efficient) were reported by our users. The results were stored and later were used as the analysis demonstration samples for training the model. 

The entire study was performed based on Polestar~\cite{Wongsuphasawat2016VoyagerEA}, which has a Tableau-liked interface and designed based on Viga-Lite and the design space $G$. The study started with a tutorial to introduce the goal and tasks of the study, and the basic operations of Polestar. Training tasks were also performed to ensure the users fully understood our goal and could generate validate analysis sequences.

As a result, 60 valid analysis sequences were collected from the pilot study (20 for each analysis task). Each of the sequences $\tilde{\zeta}_i$ is stored in the vector form, denoted as $\tilde{\zeta}_i = \{s_0, a_1, s_1, a_2, s_2, \cdots, a_{n}, s_n\}$ with $a_i$, $s_i$ indicate the action and state vector respectively.

\subsection{Policy Finding}
Based on the reward function introduced above, we are able to thread a series of charts to generate a sequence in the design space to achieve an analysis goal starting from any initial state. When the design space is modeled by an MDP, $M$, the whole sequencing process equivalent to find an optimal policy based on $M$ as described in Eq.~(\ref{eq:mdpgoal}). We employ the value-iteration based reinforcement learning algorithm~\cite{sutton1998introduction} to solve this problem. The algorithm follows the process of dynamic programming: in each iteration, it estimates the expected accumulative rewards of each state and then update the policy to ensure a better action will be executed in the next towards a state with a higher reward. 

%% file: 05-evaluation.tex
\section{Evaluation}
\label{sec:evaluation}

We evaluate our technique through a case study with an expert user and two users studies, which respectively estimate our technique in three application scenarios: (1) visualization recommendation, (2) reasoning an analysis result, and (3) making a chart design choice.

\subsection{Case Study}
A case study was conducted in a scenario of using the proposed technique for visualization recommendation. To this end, as shown in Fig.~\ref{fig:systemui}, a prototype system was developed for the case study based on Vega-Lite, D3.js~\cite{bostock2011d3}, and our chart sequencing technique. The study was performed by an expert user based on a real-world dataset. In particular, the expert was a senior PhD student with 3 years' experience in visual analysis. We used a dataset containing 406 different cars whose properties are defined by a 7-dimensional vector. 

\paragraph{\textbf{Tasks and Procedure}} 
After a brief introduction about the goal of the study and the study system, the expert was required to explore the data in the system based on the following three analysis tasks:
\begin{itemize}
    \item[\textbf{T1}] Cluster Analysis. Finding out a visualization view that clearly illustrates cluster patterns of the input data. 
    \item[\textbf{T2}] Anomaly Detection. Finding out a visualization view that clearly illustrates anomalous items in the input data.
    \item[\textbf{T3}] Correlation Analysis. Finding out a visualization view that clearly shows the correlations of the input data.
\end{itemize}

All the tasks started at the same state, i.e., a randomly picked bar chart with the X axis represents the number of \textit{Cylinders} and the Y axis represents the averaged \textit{Horsepower} as shown in Fig.~\ref{fig:teaser}. The expert was able to use any of the supported actions to modify the chart through the data mapping panel (Fig.~\ref{fig:systemui}(a)) and the chart editing panel (Fig.~\ref{fig:systemui}(b)) to finish the analysis task. The system automatically recommended actions and showed the corresponding results in the recommendation list (sorted by their rewards) for user to select (Fig.~\ref{fig:systemui}(c)) in each analysis step to guide the user to approach a selected analysis task (Fig.~\ref{fig:systemui}(d)). Each of the chart edit operations was recorded by the system and shown in the action history list (Fig.~\ref{fig:systemui}(e)).

\begin{figure}[!b]
\centering 
\includegraphics[width=\columnwidth]{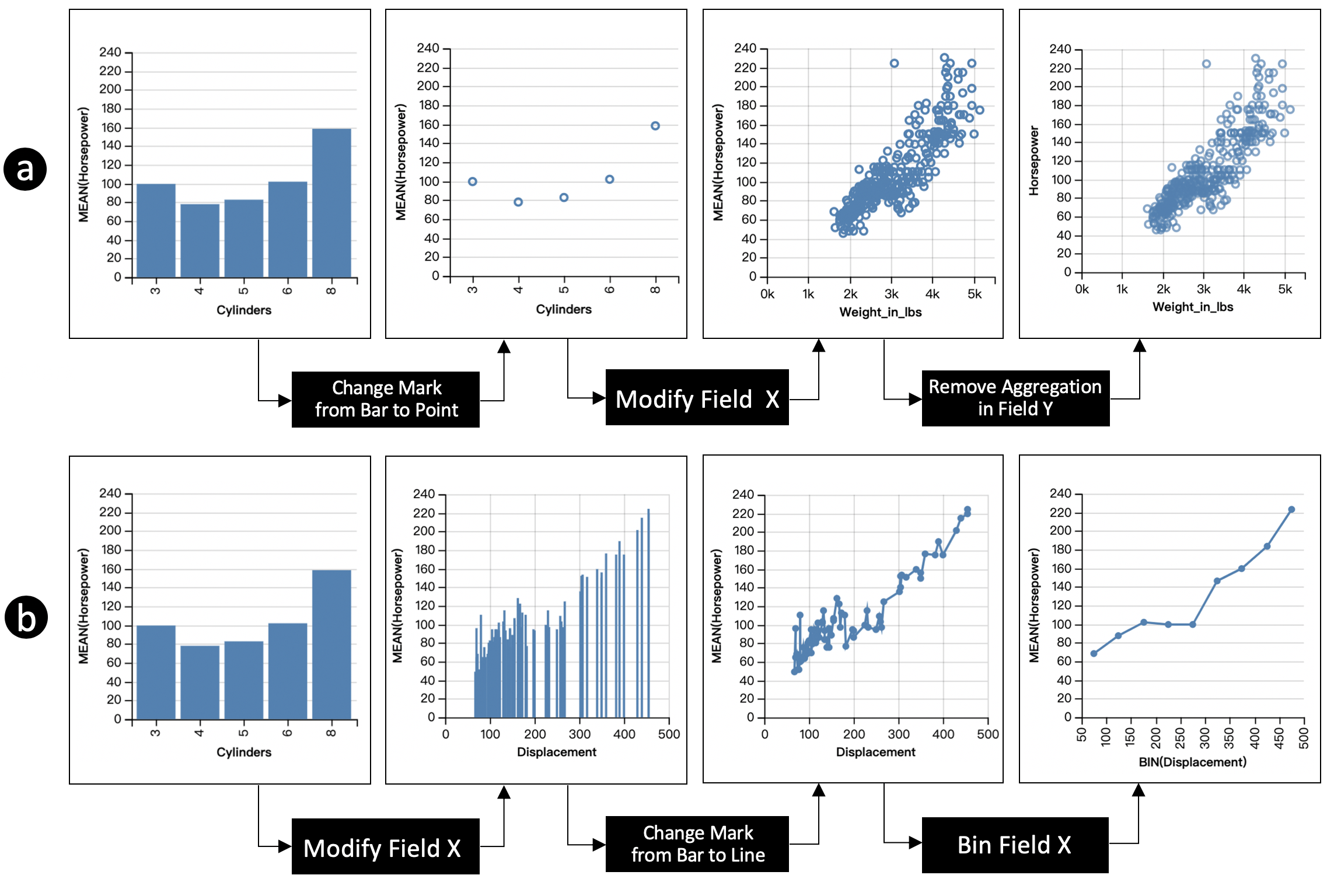}
\caption{Case study results of two analysis tasks, i.e., (a) anomaly detection and (b) correlation analysis.}
\label{fig:case}
\end{figure}

During the study, the expert was encouraged to ask questions and make comments on our system. We asked and recorded the reasons of the decisions the user made in each analysis step. A task is finished when the user reported the finding of the desired data pattern. A follow-up interview was also performed to further collect his comments on our technique. The whole study lasted for about one hour.

\paragraph{\textbf{Results}} The expert successfully finished all the tasks without meeting any trouble by following the system's recommendation. The results are shown in Fig.~\ref{fig:teaser} and Fig.~\ref{fig:case}. 

In particular, Fig.~\ref{fig:teaser} illustrates the user's exploration process of finding clusters in the data (\textbf{T1}). Starting from the initial bar chart, the user first used colors to represent the number of cylinders as he wanted to cluster cars by this property. After that, he followed the recommendations of our system by changing the chart to a scatter plot followed by a series operations on adjusting the data mappings and scales on X axis and Y axis, resulting in the last view that clearly showed several clusters. Specifically, cars with fewer cylinders (blue, orange and green dots) are clustered together as they also have a smaller horse power (Y axis) and displacement (X axis). The cars with more cylinders are also well clustered in this view.

Fig.~\ref{fig:case} illustrates the user's exploration process of finding an anomaly (\textbf{T2}) and correlation pattern (\textbf{T3}). In these two cases, the user, again, followed the system's recommendations to change the data mapping and scale on axes and change marks to switch to a proper chart type. As a result, an outlier indicates a car with a very large horse power (Y axis) but has a relatively small weight (X axis) shown in a scatter plot (Fig.~\ref{fig:case}(a)) and the displacement and the horsepower has a strong positive linear correlation (Fig.~\ref{fig:case}(b)).

\paragraph{\textbf{Feedback}} Much valuable feedback was collected during the follow-up interview. Generally, the user felt our system was very useful and the recommendation feature was ``\textit{powerful}" as it ``\textit{can help me find the answers quickly}". He also felt that the resulting sequence was ``\textit{meaningful}" and ``\textit{can help illustrate how a data pattern is detected}". He was also eager to see the techniques to be extended and used for other more complicated analysis tasks such as prediction. He also mentioned ``\textit{this is a useful feature for people who have little knowledge on data analysis, ...., but only know what they want from the data}". Despite these positive comments, after knowing more details of our technique, he also suggested ``it will be more useful if the recommendation process could take users' feedback into consideration".

\begin{figure*}[!htb]
\centering 
\includegraphics[width=2\columnwidth]{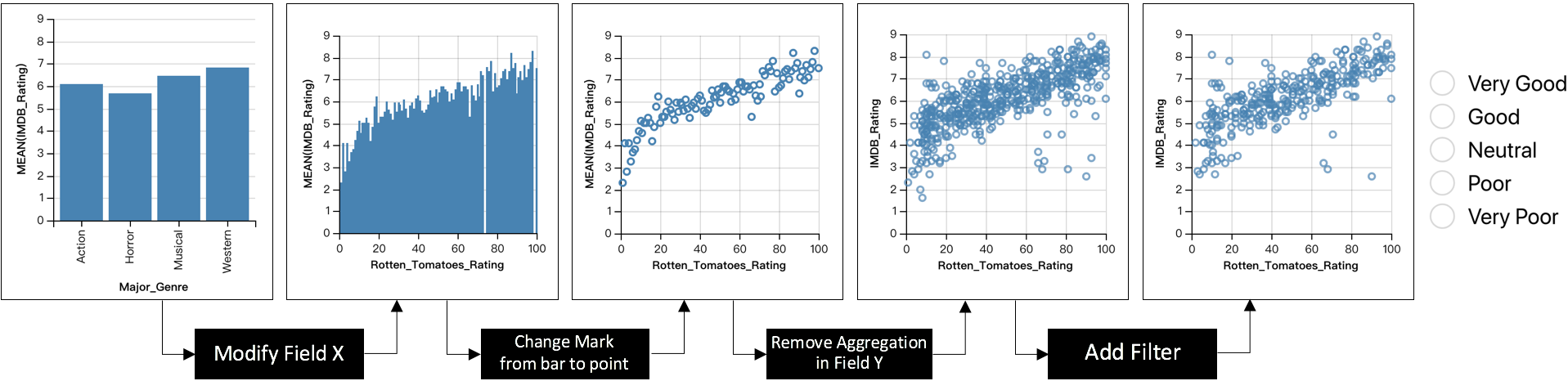}
\caption{An example chart sequence used in the user study I for participants to rate. In this study, participants were asked to rate chart sequences regarding to an analysis task based on their own preferences and experience from ``very good" to ``very poor".}
\label{fig:experiment1}
\end{figure*}

\subsection{User Studies}
We conducted two within-subject user studies with 20 participants (11 females) aged 20 to 35 (mean 25). All participants are with normal vision and reported that they have knowledge in data visualization or data analytics. All participants took part in both two user studies on two different days. The first study evaluates the effectiveness of different chart sequences, and the second one further focuses on various adjacent charts in the sequence.

\subsubsection{User Study I}
The study evaluates the effectiveness of our approach under the application scenario of sequencing charts for reasoning the analysis process. 

\paragraph{\textbf{Tasks and Procedure}}
The study was designed to measure whether the chart sequences ranked by our approach align with user preference. 
To this end, we first asked a domain expert to manually generate chart sequences as test data and then compared our ranking result of the sequences with user rating.
As a chart sequence is determined by its source and target states, we started by generating target states.
For each of the three analysis tasks, we decided a chart type and its corresponding dataset. The process of selection was supervised by the domain expert.
As a result, a line chart and two scatter plots were used for correlation analysis, anomaly detection, and cluster analysis, respectively.
The three datasets include movies, cars, and iris datasets\footnote{http://vega.github.io/voyager/}.

Each of the three visualizations was used as the target state in a chart sequence for a specific analysis task. To produce the source state in the sequence, the expert was asked to change the target state using the \textit{actions} defined in the action space $A$.  
To make the difference between the target and source states as large as possible, we required that the \textit{actions} should involve one relevant to \textit{mark} operations while the rest can be selected from options relevant to \textit{encoding} and \textit{transform} operations.
We decided upon the number of actions as four based on user requirements in a pilot study, ensuring that participants feel neither overwhelmed by information nor asking for more variations. 

\begin{figure*}[!t]
\setlength{\abovecaptionskip}{10pt}
\centering 
\includegraphics[width=0.7\textwidth]{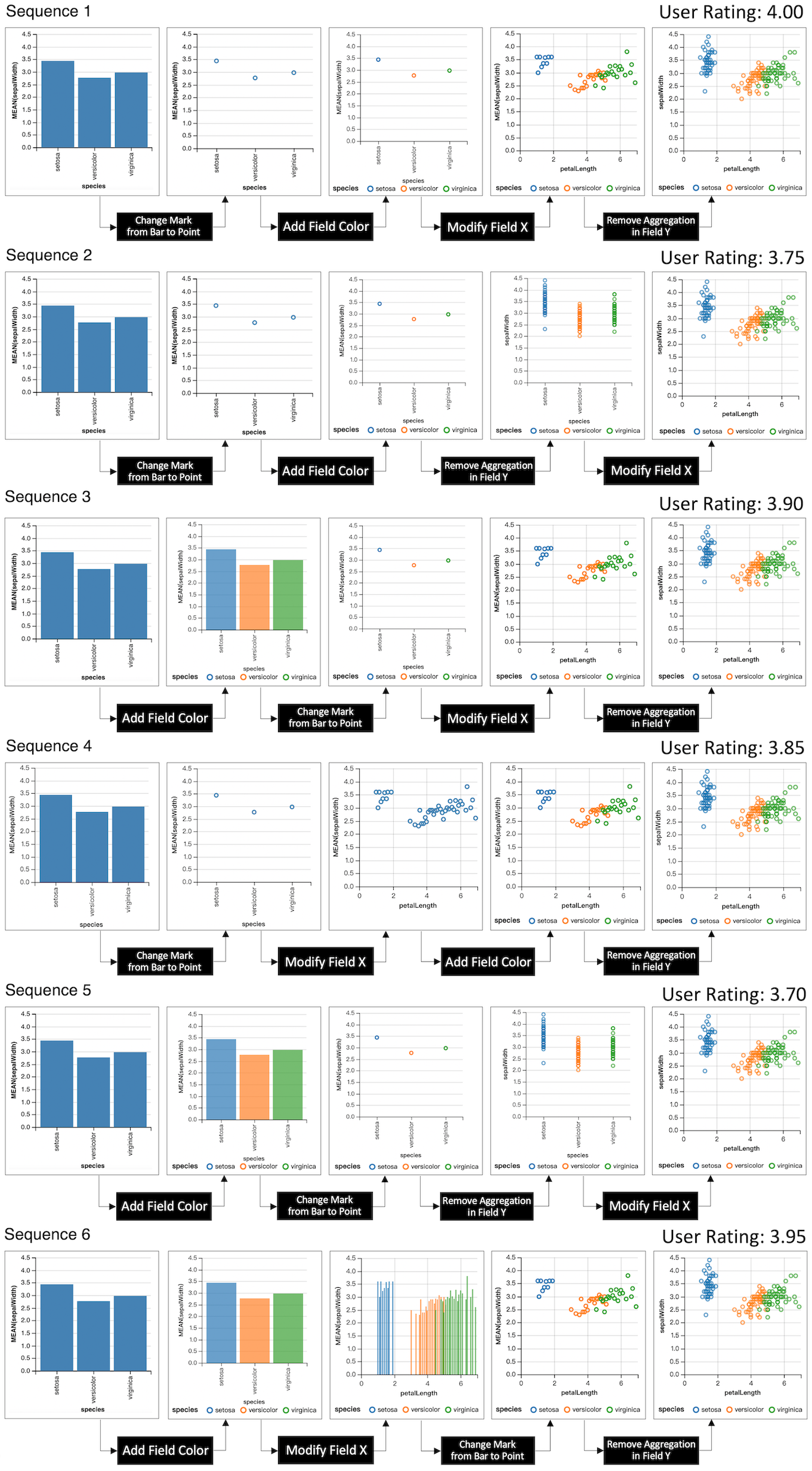}
\caption{The ranking of chart sequences used in the cluster analysis. Only the top six sequences are presented. The source state is a bar chart showing the average sepal width of different iris species. The target state is a scatter plot where the X axis represents petal length and the Y axis represents sepal width.}
\label{fig:showcase6}
\end{figure*}

Once we identified the actions of sequencing for a source-target state pair, its possible chart sequences can be enumerated via permutation. For example, given a set of four actions that transforms the source state to the target state, we can arrange the members of the action set into an order, resulting in $4!$ possible sequences.  
Thus, we generated 24 possible chart sequences for each analysis task.
For each of the three source-target state pairs, our approach ranked its 24 possible chart sequences. Fig.~\ref{fig:showcase6} shows the ranking of chart sequences used in the cluster analysis task. For demonstration purpose, only the top six sequences are displayed. The sequence ranked first uses the following actions: (1) change mark from bar to point, (2) add field color, (3) modify field X, and (4) remove aggregation in field Y.

The study consisted of three tasks, each of which based on one of the three analysis tasks: correlation analysis, anomaly detection, and cluster analysis. Before each task, we briefly described the requirements and dataset.
In each task, participants were shown a source-target state pair and its 24 possible chart sequences, as shown in Fig.~\ref{fig:experiment1}.
At the end of each task, participants were asked to rate on how well each chart sequence presents the data in a clear and logical manner~\cite{Kim2017GraphScapeAM} on a 5-point Likert scale from ``Very Good" to ``Very Poor". Participants were also encouraged to provide comments on the reason of their ratings. 
Each task lasted for about 15 minutes.
To avoid learning effects, we counterbalanced the orders of sequences as well as their assignment to the three tasks.

\paragraph{\textbf{Hypotheses}}
As we apply inverse reinforcement learning to learn a reward function for each analysis task through analysis demonstrations performed by expert users, our result should align with user preference.
Thus our hypothesis is as follows:

\begin{itemize}
  \item[\textbf{H1}] The ranking of chart sequences produced by our approach strongly correlates with participants' preference.
\end{itemize}

\paragraph{\textbf{Results}}
To measure the correlation between the ranking of chart sequences produced by our approach and by participants, we used two rank correlation statistics, Kendall's $\tau_b$ and Spearman's rank correlation coefficient. We obtained user ranking by averaging user ratings on each task and ranking the 24 chart sequences in descending order.

The results of Kendall's $\tau_b$ show a strong, positive correlation between the two rankings, with $\tau_b=0.59$, $p<0.05$. The results of Spearman's rank correlation coefficient also reveals a strong, postitive correlation between the two rankings, with $\rho=0.77,p<0.05$ (\textbf{H1 accepted}).
A majority of participants suggested that the chart sequence presenting task-related patterns in earlier stages is more preferable.  
Participants' preference for chart sequences varies in different analysis tasks.
For example, in anomaly detection task, P12 said: ``\textit{I would select the one that changes the mark type to point in the first action}''. 
P8 felt that ``\textit{the edit operations like aggregate and filter are less important in this task, I will use it in later actions}''.
In correlation analysis task, most users noted that modifying field can help understanding the relationship between columns. The actions relevant to field modification should thus be used at the beginning of sequence.
In cluster analysis task, some participants noted that encoding color should take precedence over other actions to help distinguish clusters.

\begin{figure}[!b]
\centering
\includegraphics[width=\columnwidth]{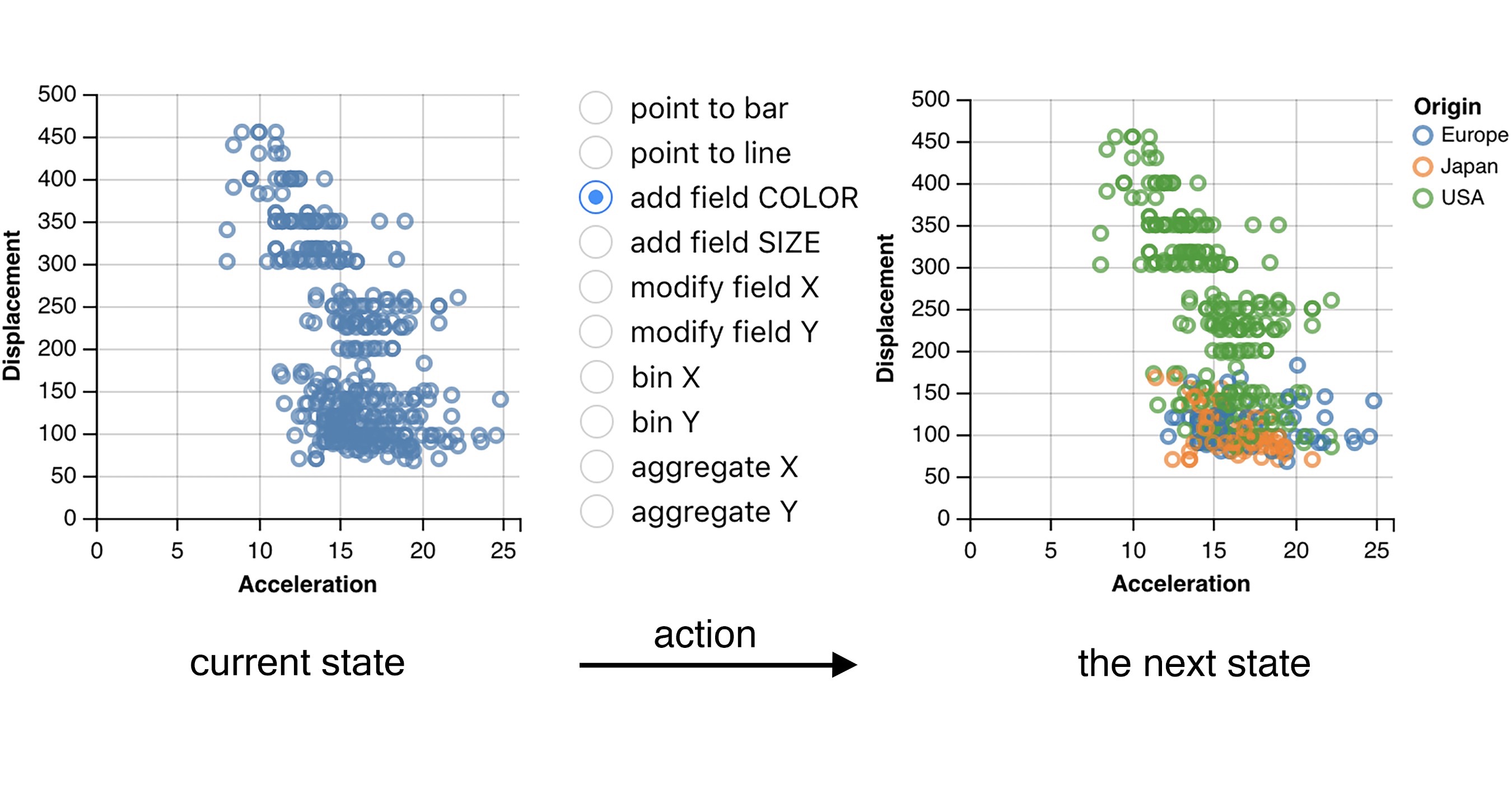}
\caption{An example design choice used in the user study II. Participants were asked to rate how possible he/she will select the edit operation as the next action with regard to an analysis task. The left visualization shows the current state while the right one shows the preview of the next state. 10 possible actions that used to transform the current state to the next state are also presented.}
\label{fig:experiment2}
\end{figure}

\subsubsection{User Study II}
The second user study evaluates the effectiveness of our approach under the application scenario of making a design choice for next chart. 

\paragraph{\textbf{Task and Procedure}}
The study was designed to measure whether the design choice recommended by our approach aligns with user preference. To this end, we first generated source states as test data and then compared our recommendation on next chart with user preference. For each analysis task, we designed three charts including a line chart, a bar chart, and a scatter plot based on iris, movies, and cars datasets, respectively. 
To avoid learning effect, we mirrored and rotated each dataset before reusing it, ensuring that participants were unable to memorize charts presented in the first study.
In total, we generated nine charts as test data for the three analysis tasks and used these nine charts as the source states.

The study procedure follows that of the first study. 
In each of the three tasks, we displayed three source states to participants one at a time, each with 10 possible next actions (Fig.~\ref{fig:experiment2}). As a participant clicks one action, he/she can preview its resulting state on the right. 
In order not to affect the participants' choice decisions, the data fields chosen for ``add/modify field" actions are random.
Participants were asked to rate how possible he/she will select the edit operation as the next action with regard to an analysis task on a 5-point Likert scale. 
Each task lasted less than 10 minutes. 
We counterbalanced the orders of charts as well as their assignment to the three tasks.

We also compared our approach with two baselines:  \textit{Task-Only}, an alternative approach by removing perception optimization from our method, and \textit{GraphScape}. 
In our method and \textit{Task-Only} condition, we ranked the 10 actions using our approach and our approach without perception optimization, respectively.
In \textit{GraphScape} condition, the 10 possible next actions were used as the input to GraphScape. We recorded the perception costs produced by GraphScape for each action and ranked the 10 actions according to their costs.

\paragraph{\textbf{Hypotheses}}
By taking both the analysis task and the factor of human cognition into consideration, our approach is able to achieve better performance compared to the baseline methods. Thus, we posed the following hypotheses:

\begin{itemize}

  \item[\textbf{H2.1}] The chart design choice recommended by our approach strongly correlates with user preference.
  \\

  \item[\textbf{H2.2}] The chart design choice recommended by our approach shows a more strongly correlation with user preference than that by \textit{GraphSpace} and \textit{Task-Only}.

\end{itemize}

\begin{table}[!t]
  \label{tab:correlation}
  \scriptsize%
  \centering%
  \begin{tabular}{%
	r%
	*{7}{c}%
	*{2}{r}%
	}
       & Kendall $\tau_b$& Spearman's coefficient\\
	    \midrule
Our method& \textbf{0.64}& \textbf{0.78}\\
Task-Only& \textbf{0.51}& \textbf{0.66}\\
GraphScape&  0.11& 0.078\\
  \midrule
  \end{tabular}%
    \caption{Correlations between the three approaches (our method, \textit{Task-Only}, \textit{GraphScape}) and user ranking. Both Kendall $\tau_b$ and Spearman's coefficient are used. Coefficients in bold are significant at $p<0.05$.}
  \label{tab:results}
\end{table}
\paragraph{\textbf{Results}}
We report statistical results and user feedback from the user study. Kendall's $\tau_b$ and Spearman's rank correlation coefficient are applied to examine if there is a strong correlation between the two rankings of design choices.

We found a strong, positive correlation between our ranking and user ranking, with $\tau_b=0.64$, $p=0.009$. Besides, the result of the Spearman's correlate coefficient is with $\rho=0.78$, $p=0.007$, which demonstrates that our ranking has a high correlation with user ranking (\textbf{H2.1} accepted). Participants also provided the motivations behind their choices. 
For example, in anomaly detection task, P4 said: ``\textit{Given that the source visualization is a scatterplot, I will add color encoding to find more patterns at first}''. 
P18 noted that``\textit{if the values are aggregated in Y axis, I would remove the aggregation to retrieve more details for anomaly detection.}''
P20 commented: ``\textit{for me, line charts may not be a good choice for revealing anomalies, so I change the line chart to a scatter plot.}''

Table~\ref{tab:results} shows that the chart design choice recommended by \textit{Task-Only} also strongly correlates with user preference, with $\tau_b=0.51, p=0.039$ and $\rho=0.66, p=0.038$. 
The coefficients between \textit{GraphScape} ranking and users ranking are $\tau_b=0.11, p=0.65$ and $\rho=0.078, p=0.829$, indicating a weak correlation. 
By comparing our approach with \textit{Task-Only} and \textit{GraphScape}, we found that our approach outperforms the two baselines (\textbf{H2.2} accepted).


%% file: 06-discussion.tex
\section{Discussion}

In this section, we discuss the limitations and implications of our current work based on the results of our evaluation. 

\subsection{Pros and Cons}

The proposed method provides many benefits besides chart recommendation, result reasoning, and design decision support, which are discussed as follows:

\begin{enumerate}
\item[{\bf P1}] \textbf{Generalization}. Based on the inverse reinforcement learning technique, our approach can be easily generalized for other analysis tasks as summarized in~\cite{Amar2005LowlevelCO} such as characterizing a distribution and finding an extremum in the data by learning the rewards based on the demonstrations performed for reaching other analysis goals. 
    
\item[{\bf P2}] \textbf{Task Oriented Chart Ranking}. Our technique assigns a reward for each state and each action given an analysis task, based on which we can rank different charts based on different tasks. We believe the resulting rank list is a useful by product that can help an analyzer who needs to use a chart to make a correct choice among various types of visualization charts.
    
\item[{\bf P3}] \textbf{Stylized Learning}. Currently, the reward function is trained by small but precise samples generated by a few experts. We believe the results can be affected by a biased training sample due to the expert's own preferences. However, sometimes, this might also be a benefit as the rewards can be trained in a personalized and stylized manner.
\end{enumerate}

Despite these benefits, we also observed several limitations of the proposed technique that need future improvement: 

\begin{enumerate}
\item[{\bf L1}] \textbf{Feedback Supporting}. The proposed technique is designed to directly generate the chart sequencing results without considering a user's feedback. It will be more useful for an explorative analysis if a chart sequencing process can also be supervised by a user's online actions. This will also make the model more robust. 

\item[{\bf L2}] \textbf{Data Supporting}. We observed the major limitation of the proposed technique is data supporting.
Our design space is highly generalized and separated from the input data. In other word, the charts are sequenced by considering their likelihood of co-occurrences in the design space based on an MDP without considering too much about underlying data to be analyzed. 
We should consider the characteristics of the selected data fields when an expert makes the decision. 
Thus a better chart sequencing results could be generated to more precisely approach an analysis goal.
However, large-scale training data should be involved due to the complexity and diversity of the datasets, which needs long-term collection work.
Besides, to avoid the high possibility of a biased model, we should also collect demonstrations from more experts with experiences to guide the training process.
This is a promising research direction that worth being studied in the future. 
\end{enumerate}

\subsection{Task Analysis vs. User Perception}
Our approach involves task analysis in the reward function to recommend chart sequences for a specific analysis task.
By comparing our approach with the baseline without perception optimization, we found that the design choices recommended by the two methods both show a strong, positive correlation with user preference (Table~\ref{tab:results}). 
The finding suggests that user perception could be a less important factor when deciding a design choice in an analysis task. Most participants selected the action that transforms the current state to the next state mainly based on the task-oriented target. 
However, in other application scenarios such as animated transition, users tend to make design choices based on their perception costs. 
Therefore, our approach should allow users to adjust the weights of task analysis and perception to generate recommendations under different application scenarios. 

%% file: 07-conclusion.tex
\section{Conclusion and Future Work}

In this paper, we introduce a novel chart sequencing method based on reinforcement learning to capture the connections between charts in the context of three major analysis tasks, including correlation analysis, anomaly detection, and cluster analysis.
We contribute (1) an approach based on reinforcement learning that seeks an optimal policy to sequencing charts in the design space to achieve a specific analysis task, and (2) a novel inverse reinforcement learning method to learn a reward function that takes into account both the analysis tasks and human cognition.
We designed and conducted two controlled user-studies to evaluate the effectiveness of our method under the application scenarios of sequencing charts for reasoning an analysis result and for making a chart design choice.
In both studies, our approach had a good performance that can match users' understanding and preference.

In the future, we plan to explore the following research directions. 
First, our approach should be extended to support chart sequencing for multiple analysis tasks. In real-world scenarios, analysts might be tasked with finishing multiple tasks at the same time.
Thus, we plan to use analysis sequences combining segments used for different tasks as sample data to train our model. 
Second, we plan to combine analysis sequences collected from expert users with the attributes of datasets (e.g., data volume, data distribution, and data type) or user feedback as training data. For example, when analyzing a large dataset, our approach can recommend aggregation operation as a possible action, which is often applied to reduce visual clutter.
Third, we should collect more analysis sequences conducted by domain experts to generate a training set of high diversity. Thus, our reinforcement learning-based model can achieve better performance.




%% file: main.bbl
\begin{thebibliography}{10}

\bibitem{plotlyfeed}
Plotly community feed.
\newblock \url{https://plot.ly/feed}.
\newblock Accessed: 2019-03-03.

\bibitem{maxent}
Principle of maximum entropy.
\newblock \url{https://en.wikipedia.org/wiki/Principle_of_maximum_entropy}.
\newblock Accessed: 2019-03-02.

\bibitem{tableaupublic}
Tableau public gallery.
\newblock \url{https://public.tableau.com/en-us/s/gallery}.
\newblock Accessed: 2019-03-03.

\bibitem{ACG14}
D.~Albers, M.~Correll, and M.~Gleicher.
\newblock Task-driven evaluation of aggregation in time series visualization.
\newblock In {\em Proceedings of the SIGCHI Conference on Human Factors in
  Computing Systems}, pp. 551--560. ACM, 2014.

\bibitem{Amar2005LowlevelCO}
R.~A. Amar, J.~R. Eagan, and J.~T. Stasko.
\newblock Low-level components of analytic activity in information
  visualization.
\newblock {\em IEEE Symposium on Information Visualization.}, pp. 111--117,
  2005.

\bibitem{TheGrandTour}
D.~Asimov.
\newblock The grand tour: a tool for viewing multidimensional data.
\newblock {\em SIAM journal on scientific and statistical computing 6.1}, pp.
  128--143, 1985.

\bibitem{Bavoil2005VisTrailsEI}
L.~Bavoil, S.~P. Callahan, C.~E. Scheidegger, H.~T. Vo, P.~Crossno, C.~T.
  Silva, and J.~Freire.
\newblock Vistrails: enabling interactive multiple-view visualizations.
\newblock {\em IEEE Visualization.}, pp. 135--142, 2005.

\bibitem{bostock2011d3}
M.~Bostock, V.~Ogievetsky, and J.~Heer.
\newblock D$^3$ data-driven documents.
\newblock {\em IEEE Transactions on Visualization and Computer Graphics},
  17(12):2301--2309, 2011.

\bibitem{Callahan2006ManagingTE}
S.~P. Callahan, J.~Freire, E.~Santos, C.~E. Scheidegger, C.~T. Silva, and H.~T.
  Vo.
\newblock Managing the evolution of dataflows with vistrails.
\newblock {\em 22nd International Conference on Data Engineering Workshops},
  pp. 71--71, 2006.

\bibitem{Dibia2018Data2VisAG}
V.~Dibia and Çagatay Demiralp.
\newblock Data2vis: Automatic generation of data visualizations using sequence
  to sequence recurrent neural networks.
\newblock {\em CoRR}, abs/1804.03126, 2018.

\bibitem{GCNF13}
M.~Gleicher, M.~Correll, C.~Nothelfer, and S.~Franconeri.
\newblock Perception of average value in multiclass scatterplots.
\newblock {\em IEEE Transactions on Visualization and Computer Graphics},
  19(12):2316--2325, 2013.

\bibitem{Gotz2009BehaviordrivenVR}
D.~Gotz and Z.~Wen.
\newblock Behavior-driven visualization recommendation.
\newblock In {\em IUI}, 2009.

\bibitem{Gotz2008CharacterizingUV}
D.~Gotz and M.~X. Zhou.
\newblock Characterizing users’ visual analytic activity for insight
  provenance.
\newblock {\em 2008 IEEE Symposium on Visual Analytics Science and Technology},
  pp. 123--130, 2008.

\bibitem{HYFC14}
L.~Harrison, F.~Yang, S.~Franconeri, and R.~Chang.
\newblock Ranking visualizations of correlation using weber's law.
\newblock {\em IEEE Transactions on Visualization and Computer Graphics},
  20(12):1943--1952, 2014.

\bibitem{Heer2008GraphicalHF}
J.~Heer, J.~D. Mackinlay, C.~Stolte, and M.~Agrawala.
\newblock Graphical histories for visualization: Supporting analysis,
  communication, and evaluation.
\newblock {\em IEEE Transactions on Visualization and Computer Graphics}, 14,
  2008.

\bibitem{Heer2007AnimatedTI}
J.~Heer and G.~G. Robertson.
\newblock Animated transitions in statistical data graphics.
\newblock {\em IEEE Transactions on Visualization and Computer Graphics},
  13:1240--1247, 2007.

\bibitem{Heer:2012}
J.~Heer and B.~Shneiderman.
\newblock Interactive dynamics for visual analysis.
\newblock {\em Communications of the ACM}, 55(4):45--54, 2012.

\bibitem{Hu2018VizNet}
K.~Hu, N.~Gaikwad, M.~Bakker, M.~Hulsebos, and E.~Zgraggen.
\newblock Viznet: Towards a large-scale visualizationlearning and benchmarking
  repository.
\newblock {\em CoRR}, 2018.

\bibitem{Hu2018VizMLAM}
K.~Z. Hu, M.~A. Bakker, S.~K.~H. Li, T.~Kraska, and C.~A. Hidalgo.
\newblock Vizml: A machine learning approach to visualization recommendation.
\newblock {\em arXiv preprint arXiv:1808.04819}, 2018.

\bibitem{Hullman2013ADU}
J.~Hullman, S.~M. Drucker, N.~H. Riche, B.~Lee, D.~Fisher, and E.~Adar.
\newblock A deeper understanding of sequence in narrative visualization.
\newblock {\em IEEE Transactions on Visualization and Computer Graphics},
  19:2406--2415, 2013.

\bibitem{Hullman2017FindingAC}
J.~Hullman, R.~Kosara, and H.~Lam.
\newblock Finding a clear path: Structuring strategies for visualization
  sequences.
\newblock {\em Comput. Graph. Forum}, 36:365--375, 2017.

\bibitem{JankunKelly2007AMA}
T.~J. Jankun-Kelly, K.-L. Ma, and M.~Gertz.
\newblock A model and framework for visualization exploration.
\newblock {\em IEEE Transactions on Visualization and Computer Graphics}, 13,
  2007.

\bibitem{KH16}
M.~Kay and J.~Heer.
\newblock Beyond weber's law: A second look at ranking visualizations of
  correlation.
\newblock {\em IEEE Transactions on Visualization and Computer Graphics},
  22(1):469--478, 2016.

\bibitem{Kim2018AssessingEO}
Y.~Kim and J.~Heer.
\newblock Assessing effects of task and data distribution on the effectiveness
  of visual encodings.
\newblock {\em Comput. Graph. Forum}, 37:157--167, 2018.

\bibitem{Kim2017GraphScapeAM}
Y.~Kim, K.~Wongsuphasawat, J.~Hullman, and J.~Heer.
\newblock Graphscape: A model for automated reasoning about visualization
  similarity and sequencing.
\newblock In {\em Proceedings of the CHI Conference on Human Factors in
  Computing Systems}, 2017.

\bibitem{Luo2018DeepEyeTA}
Y.~Luo, X.~Qin, N.~Tang, and G.~Li.
\newblock Deepeye: Towards automatic data visualization.
\newblock {\em 2018 IEEE 34th International Conference on Data Engineering
  (ICDE)}, pp. 101--112, 2018.

\bibitem{Ma1999ImageGN}
K.-L. Ma.
\newblock Image graphs-a novel approach to visual data exploration.
\newblock {\em Proceedings Visualization (Cat. No.99CB37067)}, pp. 81--88,
  1999.

\bibitem{Mackinlay1986AutomatingTD}
J.~D. Mackinlay.
\newblock Automating the design of graphical presentations of relational
  information.
\newblock {\em ACM Trans. Graph.}, 5:110--141, 1986.

\bibitem{Mackinlay2007ShowMA}
J.~D. Mackinlay, P.~Hanrahan, and C.~Stolte.
\newblock Show me: Automatic presentation for visual analysis.
\newblock {\em IEEE Transactions on Visualization and Computer Graphics}, 13,
  2007.

\bibitem{Moritz2018FormalizingVD}
D.~Moritz, C.~Wang, G.~L. Nelson, H.~Lin, A.~M. Smith, B.~Howe, and J.~Heer.
\newblock Formalizing visualization design knowledge as constraints: Actionable
  and extensible models in draco.
\newblock {\em IEEE Transactions on Visualization and Computer Graphics},
  25:438--448, 2018.

\bibitem{Qu2018KeepingMV}
Z.~Qu and J.~Hullman.
\newblock Keeping multiple views consistent: Constraints, validations, and
  exceptions in visualization authoring.
\newblock {\em IEEE Transactions on Visualization and Computer Graphics},
  24:468--477, 2018.

\bibitem{Saket2018TaskBasedEO}
B.~Saket, A.~Endert, and C.~Demiralp.
\newblock Task-based effectiveness of basic visualizations.
\newblock {\em IEEE Transactions on Visualization and Computer Graphics}, 2018.

\bibitem{Satyanarayan2017VegaLiteAG}
A.~Satyanarayan, D.~Moritz, K.~Wongsuphasawat, and J.~Heer.
\newblock Vega-lite: A grammar of interactive graphics.
\newblock {\em IEEE Transactions on Visualization and Computer Graphics},
  23:341--350, 2017.

\bibitem{Scheidegger2007QueryingAC}
C.~E. Scheidegger, H.~T. Vo, D.~Koop, J.~Freire, and C.~T. Silva.
\newblock Querying and creating visualizations by analogy.
\newblock {\em IEEE Transactions on Visualization and Computer Graphics},
  13:1560--1567, 2007.

\bibitem{Segel2010NarrativeVT}
E.~Segel and J.~Heer.
\newblock Narrative visualization: Telling stories with data.
\newblock {\em IEEE Transactions on Visualization and Computer Graphics},
  16:1139--1148, 2010.

\bibitem{sutton1998introduction}
R.~S. Sutton, A.~G. Barto, et~al.
\newblock {\em Introduction to reinforcement learning}, vol. 135.
\newblock MIT press Cambridge, 1998.

\bibitem{SHGF16}
D.~A. Szafir, S.~Haroz, M.~Gleicher, and S.~Franconeri.
\newblock Four types of ensemble coding in data visualizations.
\newblock {\em Journal of vision}, 16(5):11--11, 2016.

\bibitem{Wongsuphasawat2016VoyagerEA}
K.~Wongsuphasawat, D.~Moritz, A.~Anand, J.~D. Mackinlay, B.~Howe, and J.~Heer.
\newblock Voyager: Exploratory analysis via faceted browsing of visualization
  recommendations.
\newblock {\em IEEE Transactions on Visualization and Computer Graphics},
  22:649--658, 2016.

\bibitem{ziebart2008maximum}
B.~D. Ziebart, A.~L. Maas, J.~A. Bagnell, and A.~K. Dey.
\newblock Maximum entropy inverse reinforcement learning.
\newblock In {\em AAAI}, vol.~8, pp. 1433--1438. Chicago, IL, USA, 2008.

\end{thebibliography}
